\documentclass[11pt]{article}
\usepackage{calc}
\usepackage{rotating}
\usepackage[english]{babel}
\usepackage{graphicx}
\usepackage{subfig}
\usepackage{float}
\usepackage{amsmath}
\usepackage{amssymb}
\usepackage{amsthm}
\usepackage{latexsym}
\usepackage{dcolumn}
\usepackage{geometry}
\usepackage{color}
\usepackage{hyperref}
\usepackage{mciteplus}
\usepackage{slashed}
\usepackage{multirow}
\def\gtwid{\mathrel{\raise.3ex\hbox{$>$\kern-.75em\lower1ex\hbox{$\sim
$}}}}
\def\vio{\mathrel{\hbox{$E$\kern-.60em\hbox{$/
$}}}}
\newcommand{\nn}{\nonumber}

\newcommand{\hobs}{H_{\rm obs}}

\newcommand{\HiS}{HiggsSignals v1.3.2}
\newcommand{\HiB}{HiggsBounds v4.2.0}

\begin{document}

\begin{flushright}
APCTP-PRE2015-011\\
%SHEP-15-**\\
\end{flushright}
\vspace*{2.0cm}
\begin{center}
{\Large \bf {Two Higgs bosons near 125\,GeV in the complex NMSSM \\[0.45cm]
and the LHC Run-I data}}\\
\vspace*{0.8cm}
{\large Stefano Moretti$^{a}$ and Shoaib Munir$^{b}$ \\[2ex]
{\it \small $^a$School of Physics \& Astronomy, \\
University of Southampton, Southampton SO17 1BJ, UK \\[0.25cm]
$^b$Asia Pacific Center for Theoretical Physics, San 31, Hyoja-dong, \\
  Nam-gu, Pohang 790-784, Republic of Korea}
\bigskip \\
\url{s.moretti@soton.ac.uk, s.munir@apctp.org}
}
\end{center}
\vspace*{0.4cm}

\begin{abstract}
\noindent
We analyse the impact of explicit CP-violation in the Higgs sector of
the Next-to-Minimal Supersymmetric Standard Model (NMSSM) on
its consistency with the Higgs boson data from the Large Hadron
Collider (LHC). Through detailed scans of the parameter space of the 
complex NMSSM for certain fixed values of one of its CP-violating (CPV)
phases, we obtain a large number of points corresponding to five 
phenomenologically relevant scenarios containing $\sim 125$\,GeV Higgs boson(s). 
We focus, in particular, on the scenarios where the
visible peaks in the experimental samples can actually be explained by
two nearly mass-degenerate neutral Higgs boson states. We find that 
some points corresponding to these scenarios give an overall slightly improved fit 
to the data, more so for non-zero values of the CPV phase, compared to
the scenarios containing a single Higgs boson near 125\,GeV. 
\end{abstract}

\newpage
\section{\label{intro}Introduction}

The Higgs sector of the NMSSM~\cite{Fayet:1974pd,*Ellis:1988er,*Durand:1988rg,*Drees:1988fc} (see,
e.g.,~\cite{Ellwanger:2009dp,Maniatis:2009re} for reviews) contains
two additional neutral mass eigenstates besides the three of the
Minimal Supersymmetric Standard Model (MSSM). This is due to
the presence of a Higgs singlet
superfield besides the two doublet superfields of the MSSM. 
When all the parameters in the Higgs and sfermion
sectors of the NMSSM are real, one of these new Higgs states is a scalar and the other a
pseudoscalar. Hence, in total three scalars, $H_{1,2,3}$, and two
pseudoscalars, $A_{1,2}$, make up the neutral Higgs boson content
of the model. This extended Higgs sector of the NMSSM boasts some unique
phenomenological possibilities, which are either precluded or
experimentally ruled out in the MSSM. For example, in the NMSSM either of the two 
lightest CP-even Higgs bosons, $H_1$ or  $H_2$, can play the role of
the $\sim 125$\,GeV Standard Model (SM)-like Higgs boson, $\hobs$, observed at the
LHC~\cite{Aad:2012tfa,Chatrchyan:2012ufa,Chatrchyan:2013lba}. 

Of particular interest in the NMSSM is the possibility that the
SM-like Higgs boson can obtain a large tree-level mass in a {\it natural}
way, i.e., without requiring large
radiative corrections from the supersymmetric sectors. This happens
in a specific region of the parameter space, which we refer to as
the natural NMSSM, where there is a significant singlet-doublet mixing
and the $\hobs$ is typically $H_2$.
%The scenario where the $H_2$ can be identified with the $\hobs$ is
%realised in a particular parameter space region of the NMSSM, where it
%gains a mass near 125\,GeV {\it naturally}, i.e., without requiring large
%radiative corrections from the supersymmetric sectors. 
This scenario was used to
explain~\cite{Ellwanger:2011aa,*King:2012is,*Cao:2012fz} the
enhancement in the $\hobs
\rightarrow \gamma\gamma$ channel in the early
LHC data. However, when the singlet-doublet mixing is too large, the properties
of $H_2$ can deviate appreciably from an exact SM-like behaviour,
resulting in a reduction of its fermionic partial decay widths.
An alternative possibility in a very similar parameter space region is
that of both $H_1$ and $H_2$ simultaneously having masses near
125\,GeV~\cite{Gunion:2012gc,King:2012tr,*Gherghetta:2012gb,*Wu:2015nba}. In that case, the observed excess at the
LHC could actually be due to a superposition of these two states, when
their individual
signal peaks cannot be resolved separately. One of these two Higgs
bosons, typically $H_1$, is the singlet-like neutral state. 
Moreover, in~\cite{Munir:2013wka} it was noted that 
the lighter of the two pseudoscalars, $A_1$, when it is singlet-like, 
could also be nearly mass-degenerate with a SM-like
$H_1$ near 125\,GeV, instead of or even along with the $H_2$. However, such a
pseudoscalar can only contribute visibly to the measured signal
strength near 125\,GeV if it is produced in association with a $b\bar{b}$ pair.
	
One of the most important yet unresolved issues in particle
physics is that of the observed matter-antimatter asymmetry in the 
universe. A plausible explanation for this asymmetry is
 electroweak (EW) baryogenesis~\cite{Cohen:1993nk,*Quiros:1994dr,*Rubakov:1996vz*Trodden:1998ym}. The necessary
conditions for successful EW baryogenesis 
include the following~\cite{Sakharov:1967dj}: (1) baryon number violation, 
(2) CP-violation and (3) departure from equilibrium at the critical 
temperature of the EW symmetry breaking (EWSB) phase transition, implying 
that it is strongly first order. In the SM, a strongly first order
EW phase transition is not possible given the
measured mass of the Higgs boson at the LHC. Besides, the only
source of CP-violation in the SM, the Cabibbo-Kobayashi-Maskawa 
matrix, is insufficient. 
Therefore, beyond the SM, a variety of sources of CP-violation have 
been proposed in the literature (for a review,
see~\cite{Ibrahim:2007fb} and references therein). 
In the context of supersymmetry (SUSY), a strongly first 
order phase transition is possible in the MSSM only if the lightest stop has 
a mass below that of the top quark. This possibility has now been 
ruled out by SUSY searches at the 
LHC~\cite{Cohen:2012zza,*Curtin:2012aa,*Carena:2012np}. 
Also, the MSSM Higgs sector does not violate CP at the
tree level but does so only at higher 
orders~\cite{Pilaftsis:1998pe,*Pilaftsis:1998dd,Pilaftsis:1999qt,*Carena:2000yi,
Choi:2000wz,*Carena:2001fw,*Carena:2002bb,*Choi:2004kq,
*Frank:2006yh,*Heinemeyer:2007aq}. The
CPV phases, transmitted radiatively
to the Higgs sector via couplings to the sfermions, are tightly
constrained by the measurements of fermion electric dipole moments 
(EDMs)~\cite{Baker:2006ts,*Commins:2007zz,*Griffith:2009zz}. However,
these EDM constraints can be relaxed under certain 
conditions~\cite{Pilaftsis:1999qt,*Carena:2000yi,Abel:2001vy,Haba:1996bg,*Ibrahim:1998je,*Boz:2005sf,*Ellis:2008zy,*Li:2010ax}.

The NMSSM has been shown to accommodate a strongly first order EW
phase transition without a light stop~\cite{Huber:2000mg,*Huber:2006ma,*Kanemura:2011fy,*Cheung:2012pg,*Huang:2014ifa,*Bi:2015qva}. Additionally, in this model, CP-violation can
be invoked explicitly in the Higgs sector even at the tree level by
assuming the Higgs self-couplings, $\lambda$ and $\kappa$, to be
complex. Beyond the Born approximation, the phase of 
the SUSY-breaking Higgs-sfermion-sfermion couplings, $A_{\tilde{f}}$, 
where $f$ denotes a SM fermion, is also induced in the Higgs sector, 
as in the MSSM.
In the presence of the associated complex phases, the five neutral Higgs 
bosons are CP-indefinite states, due to the mixing between 
the scalar and pseudoscalar interaction eigenstates.
CPV phases can therefore influence the phenomenology 
of the NMSSM Higgs bosons by, e.g., modifying their mass spectrum as well as their 
production and decay rates~\cite{Moretti:2013lya}, similarly to the
MSSM~\cite{Demir:1999hj,*Dedes:1999sj,*Dedes:1999zh,*Kane:2000aq,*Arhrib:2001pg,
*Choi:2001pg,*Choi:2002zp,*Ellis:2004fs,*Hesselbach:2009gw,*Fritzsche:2011nr,*Chakraborty:2013si}.
The impact of these phases in the complex NMSSM (cNMSSM), i.e., the
CPV NMSSM, on the necessary
conditions for successful EW phase transition was also studied
some time ago~\cite{Funakubo:2005pu}. The consistency of scenarios 
yielding the correct baryon asymmetry with the LHC Higgs boson data 
still remains to be studied in depth, though. However, even 
leaving aside these considerations, the possibly distinct
phenomenological scenarios that the cNMSSM can yield make
it a particularly interesting model for exploration at the Run-II of the LHC. 

The cNMSSM has therefore been the subject of several studies 
recently and, in particular, some important theoretical developments
have been made in the model. The dominant 1-loop corrections
to the neutral Higgs sector from the (s)quark and gauge sectors were 
studied in\,\cite{Ham:2001wt,*Funakubo:2004ka,Cheung:2010ba,*Cheung:2011wn}, 
in the renormalisation group equations-improved effective potential
approach. The corrections from the gaugino sector were included
in~\cite{Munir:2013dya} and, more inclusively, recently in~\cite{Domingo:2015qaa}. 
In the Feynman diagrammatic approach, the
complete 1-loop Higgs mass matrix was derived in~\cite{Graf:2012hh} and
the $\mathcal{O}(\alpha_t \alpha_s)$ contributions to it were calculated
in~\cite{Muhlleitner:2014vsa}. As far as the phenomenology of the
Higgs bosons in the cNMSSM is concerned, the consistency of several
CPV scenarios with the early results on the $\hobs$ from 
the LHC data was studied in detail in~\cite{Graf:2012hh,Moretti:2013lya}. 
Another distinct phenomenological scenario, possible only
for non-zero CPV phases, has also been studied in~\cite{Munir:2013dya}. 

The CMS and ATLAS collaborations have recently updated their
measurements of the $\hobs$ signal rates in the $\tau^+\tau^-$ 
and $b\bar{b}$ channels~\cite{Khachatryan:2014jba,ATLAS-CONF-2014-009}. 
%The fact that these rates also
% tend to favour a SM-like $\hobs$ is increasingly jeopardising the
%above mentioned natural scenario with a $125$\,GeV $H_2$ in the
%real NMSSM (rNMSSM), i.e., the CPC NMSSM. 
The fact that these rates also
 tend to favour a SM-like $\hobs$ is increasingly jeopardising the
above mentioned natural NMSSM scenario with large singlet-doublet mixing
but only with one Higgs boson, either $H_1$ or $H_2$, around 125\,GeV.
This makes the scenario with both $H_1$ and $H_2$
contributing to the observed $\sim 125$\,GeV signal all the more
important, since it may potentially satisfy better the current Higgs
boson data while still leaving plenty of room for new physics. In case of 
the cNMSSM, since the five neutral Higgs bosons are CP-mixed states, 
the scenario with mass-degenerate $H_1$ and $H_2$ can entail both the
corresponding possibilities in the real NMSSM (rNMSSM), i.e., 
mass-degenerate $H_1,\,H_2$ or $H_1/H_2,\,A_1$. 

In this study we therefore analyse and 
compare the prospects for scenarios with two mass-degenerate Higgs
bosons against those with a single Higgs boson near 125\,GeV in the
% natural 
$Z_3$-invariant cNMSSM. We perform scans of the 
relevant parameter 
space~\cite{Gunion:2012gc} of the model using the 
public program NMSSMCALC~\cite{Baglio:2013iia} to search for all possible $\sim 125$\,GeV Higgs boson scenarios, with
the CPV phase of the coupling $\kappa$ set to five different values, including
$0^\circ$ - the rNMSSM limit - in each case. 
The condition for mass-degeneracy between two Higgs bosons is imposed
by requiring them to lie within 2.5\,GeV of each other, which is
consistent with the current mass resolution of the
LHC~\cite{Aad:2015zhl}, taking into account the
uncertainties in the theoretical mass prediction. We then use fits to
the Higgs boson data from the LHC Run-I, both with $\sqrt{s}=7$\,TeV 
and $\sqrt{s}=8$\,TeV, as well as from the Tevatron, performed using
 the program HiggsSignals~\cite{Bechtle:2013xfa}, as 
the sole criterion for comparing the present likelihood of each of these scenarios.
We also discuss how these mass-degenerate Higgs bosons 
can be identified at the LHC based on 
the signal rate double ratios introduced in~\cite{Gunion:2012he}.

The article is organised as follows. In the next section we will 
briefly revisit the Higgs sector of the cNMSSM. In
section \ref{sec:method} we will provide details of our numerical scans and our
procedure for fitting the model predictions for the Higgs
boson(s) to the LHC data. In section \ref{sec:results} we
will discuss the results of our analysis and in section
\ref{sec:concl} we will present our conclusions.

\section{\label{sec:model}The Higgs sector of the cNMSSM}

The NMSSM contains a singlet Higgs
superfield, $\widehat{S}$, besides the two $SU(2)_L$ doublet
superfields,
 \begin{eqnarray}
\widehat{H}_u = \left(\begin{array}{c} \widehat{H}_u^+ \\ \widehat{H}_u^0
\end{array}\right)\,,\;
\widehat{H}_d = \left(\begin{array}{c} \widehat{H}_d^0 \\ \widehat{H}_d^-
\end{array}\right)\,,
\end{eqnarray}
of the MSSM. The superpotential of the NMSSM is written as
\begin{equation}
\label{eq:superpot}
W_{\rm NMSSM}\ =\ {\rm MSSM\;Yukawa\;terms} \: +\: 
\lambda \widehat{S} \widehat{H}_u \widehat{H}_d \: + \:
\frac{\kappa}{3}\ \widehat{S}^3\,,
\end{equation}
where $\lambda$ and $\kappa$ are dimensionless Yukawa couplings. This
superpotential is scale invariant, since the term $\mu \widehat{H}_u \widehat{H}_d$
appearing in the MSSM superpotential has been removed by imposing a discrete $Z_3$
symmetry. In this model, an effective $\mu$-term, $\mu_{\rm eff} =
\lambda s$, is instead generated when the singlet field
acquires a vacuum expectation value (VEV), $s$, which is naturally of the order of the
SUSY-breaking scale.

The tree-level Higgs potential of the NMSSM, obtained from the superpotential in
eq.\,(\ref{eq:superpot}), is written in terms of the neutral scalar
components of the Higgs superfields, $H_u$, $H_d$, and $S$, as
\begin{eqnarray}
\label{eq:Higgspot}
V_0 & = & \left|\lambda \left(H_u^+ H_d^- - H_u^0
H_d^0\right) + \kappa S^2 \right|^2 \nn \\
&&+\left(m_{H_u}^2 + \left|\lambda S\right|^2\right) 
\left(\left|H_u^0\right|^2 + \left|H_u^+\right|^2\right) 
+\left(m_{H_d}^2 + \left|\lambda S\right|^2\right) 
\left(\left|H_d^0\right|^2 + \left|H_d^-\right|^2\right) \nn \\
&&+\frac{g^2}{4}\left(\left|H_u^0\right|^2 +
\left|H_u^+\right|^2 - \left|H_d^0\right|^2 -
\left|H_d^-\right|^2\right)^2
+\frac{g_2^2}{2}\left|H_u^+ H_d^{0*} + H_u^0 H_d^{-*}\right|^2\nn \\
&&+m_{S}^2 |S|^2
+\big( \lambda A_\lambda \left(H_u^+ H_d^- - H_u^0 H_d^0\right) S + 
\frac{1}{3} \kappa A_\kappa\, S^3  + \mathrm{h.c.}\big)\,,
\end{eqnarray}
where $g^2\equiv \frac{g_1^2 + g_2^2}{2}$, with $g_1$ and $g_2$ being
the $U(1)_Y$ and $SU(2)_L$ gauge couplings, respectively, and $A_\lambda$
and $A_\kappa$ are the soft SUSY-breaking Higgs trilinear couplings.  
The scalar fields $H_u$, $H_d$ and $S$ are developed around their 
respective VEVs, $v_u$, $v_d$ and $s$, as %
\begin{eqnarray}
H_d^0&=&
\hphantom{e^{i\theta}}\,
\left(
\begin{array}{c}
\frac{1}{\sqrt{2}}\,(v_d+H_{dR}+iH_{dI}) \\
H_d^-
\end{array}
\right)\,, \nonumber \\[1mm]
H_u^0&=&
e^{i\theta}\,\left(
\begin{array}{c}
H_u^+\\
\frac{1}{\sqrt{2}}\,(v_u+H_{uR}+i H_{uI})
\end{array}
\right)\,, \\[1mm]
S&=&\frac{e^{i\varphi}}{\sqrt{2}}\,(s+S_R+iS_I)\,. \nonumber
\label{eq:higgsparam}
\end{eqnarray}

The Higgs coupling parameters appearing in the potential in
eq.\,(\ref{eq:Higgspot}) can very well be complex, implying $\lambda\equiv
|\lambda|e^{i\phi_\lambda}$, $\kappa\equiv
|\kappa|e^{i\phi_\kappa}$, $A_\lambda \equiv |A_\lambda|e^{i\phi_{A_\lambda}}$
and $A_\kappa \equiv |A_\kappa|e^{i\phi_{A_\kappa}}$. As a result, the
$V_0$, evaluated at the vacuum, contains the phase combinations
 \begin{equation}
\phi^\prime_\lambda \equiv \phi_\lambda+\theta+\varphi\,,\;
\phi^\prime_\kappa \equiv \phi_\kappa+3\varphi\,,\;\phi^\prime_\lambda+\phi_{A_\lambda}\;{\rm and}\;\phi^\prime_\kappa+\phi_{A_\kappa}\,.
\end{equation}
 For correct EWSB, the Higgs potential should have a minimum at 
non-vanishing $v_u$, $v_d$ and $s$, which is ensured by requiring 
\begin{eqnarray}
\bigg\langle \frac{\delta V_0}{\delta \Phi} \bigg\rangle =0\,~~~{\rm for}~~~\Phi=H_{dR},\,H_{uR},\,S_R,\,H_{dI},\,H_{uI},\,S_I\,.
\end{eqnarray}
Through the above minimisation conditions the phase combinations 
 $\phi^\prime_\lambda+\phi_{A_\lambda}$ and
 $\phi^\prime_\kappa+\phi_{A_\kappa}$ can be determined up to a 
twofold ambiguity by $\phi^\prime_\lambda-\phi^\prime_\kappa$. Thus,
$\phi^\prime_\lambda-\phi^\prime_\kappa$ is the only physical CP phase 
appearing in the NMSSM Higgs sector at the tree level. Also, using
these conditions, the soft mass
parameters $m^2_{H_u}$, $m^2_{H_d}$ and $m^2_S$ can be traded for
the corresponding Higgs field VEVs.

The neutral Higgs mass matrix is obtained by taking the second derivative of
 the $V_0$ evaluated at the vacuum. 
% has the form 
%\begin{eqnarray}
%\label{eq:mhiggs}
%\hspace{-1.0cm}
%\mathcal{M}^2_0 =
%\left(
%        \begin{array}{cc}
%\mathcal{M}_S^2 & \mathcal{M}^{2}_{SP} \\
%(\mathcal{M}^{2}_{SP})^T & \mathcal{M}_P^2  
%                \end{array}
%\right)\,.
%\end{eqnarray}
This $5\times 5$ matrix, ${\cal M}_0^2$, in the 
${\bf H}^T=(H_{dR},\,H_{uR},\,S_R,\,H_I,\,S_I)$ basis, from which the
 massless Nambu-Goldstone mode has been rotated away, 
can be diagonalised using an orthogonal matrix, $O$, as 
$O^T{\mathcal{M}}_0^2O={\rm diag}
(m^2_{H_1}\;m^2_{H_2}\;m^2_{H_3}\;m^2_{H_4}\;m^2_{H_5})$. This yields the
physical tree-level masses corresponding to the five mass eigenstates,
\begin{eqnarray}
\label{eq:rot66}
(H_1,\,H_2,\,H_3,\,H_4,\,H_5)_a^T = O_{ai}\,
(H_{dR},\,H_{uR},\,S_R,\,H_I,\,S_I)^T_i \,,
\end{eqnarray} 
 such that $m^2_{H_1}\leq m^2_{H_2}\leq m^2_{H_3} \leq m^2_{H_4} \leq m^2_{H_5}$.
The elements, $O_{ai}$, of the mixing matrix then govern the couplings of the
Higgs bosons to all the particles in the model.

The tree-level Higgs mass matrix is subject to
higher order corrections from the SM
fermions, from the gauge and 
chargino/neutralino sectors and the Higgs sector itself, 
as well as from the sfermion sector, in the case
of which they are dominated by the stop contributions. 
Upon the inclusion of these corrections, $\Delta\mathcal{M}^2$, 
the Higgs mass matrix gets modified, so that
\begin{eqnarray}
\label{eq:mhiggs-nor}
\mathcal{M}^2_H = \mathcal{M}^2_0 + \Delta\mathcal{M}^2\,.
\end{eqnarray}
Explicit expressions for $\mathcal{M}^2_0$ as well as
for $\Delta\mathcal{M}^2$ can be found 
in~\cite{Graf:2012hh,Munir:2013dya,Domingo:2015qaa}.
Thus, beyond the Born approximation, the CPV phases of the gaugino
mass parameters, $M_{1,2}$, and of $A_{\tilde{f}}$ are also radiatively 
induced in the Higgs sector of the NMSSM.

Therefore, when studying the phenomenology of the Higgs bosons, one needs to 
take into account also the parameters from the other sectors of the
model. However, the most general NMSSM contains more 
than 130 parameters at the EW scale.
Assuming the matrices for the sfermion masses and for the
trilinear scalar couplings to be diagonal considerably reduces the number of free
parameters. One can further exploit the fact, mentioned above, that
the corrections to
the Higgs boson masses from the sfermions are largely dominated by the
stop sector. For our numerical analysis in the following sections, we will
thus impose the following supergravity-inspired universality conditions on
the model parameters at the EW scale:
\begin{gather}
M_0 \equiv M_{Q_{1,2,3}} = M_{U_{1,2,3}} = M_{D_{1,2,3}} =
M_{L_{1,2,3}} = M_{E_{1,2,3}}\,, \nonumber \\
M_{1/2} \equiv 2M_1 = M_2 =  \frac{1}{3} M_3\,, \\
A_0 \equiv A_{\tilde{t}} = A_{\tilde{b}} = A_{\tilde{\tau}}\,, \nonumber
\end{gather}
 where $M^2_{Q_{1,2,3}},\,M^2_{U_{1,2,3}},\,M^2_{D_{1,2,3}}
 ,\,M^2_{L_{1,2,3}}$ and $M^2_{E_{1,2,3}}$ are the squared soft masses of the
 sfermions, $M_{1,2,3}$ those of the gauginos and $
A_{\tilde{t},\tilde{b},\tilde{\tau}}$ the soft trilinear couplings. 
Altogether, the input parameters of the cNMSSM then include
\begin{center}
$M_0$\,, $|M_{1/2}|$\,, $|A_0|$, $\tan\beta \;(\equiv v_u/v_d)$\,,
$|\lambda|$\,, $|\kappa|$\,, $\mu_{\rm eff}$\,, $|A_\lambda|$\,,
$|A_\kappa|$\,, $\theta_{1/2}$\,, $\theta_{\tilde{f}}$\,,
$\phi^\prime_\lambda$ and $\phi^\prime_\kappa$\,,
\end{center}
where $\theta_{1/2}$ and $\theta_{\tilde{f}}$ are the phases of the
unified parameters $M_{1/2}$ and $A_0$, respectively.  
%Finally, we will fix $\rm{sign}[cos(\phlam + \phi_{A_\lambda})] =
%\rm{sign}[cos(\phkap + \phi_{A_\kappa})] = +1$. 

\section{\label{sec:method} Numerical analysis}

As noted in the Introduction, non-zero CPV phases can
modify appreciably the masses and decay
widths of the neutral Higgs bosons compared to the CP-conserving case
for a given set of the remaining free parameters. In the case of the $\hobs$
candidate in the model, whether $H_1$ or $H_2$ or even $H_3$, the CPV
phases are thus strongly constrained by the LHC mass and signal rate 
measurements. This was analysed in detail in~\cite{Moretti:2013lya}, 
where the scenarios with mass-degenerate Higgs bosons were, 
however, not taken into account. In the present study we
thus test whether the said modifications in the Higgs boson
properties with non-zero values of the phase $\phi^\prime_\kappa$ 
%(with $\varphi = 0^\circ$, so that $\phi^\prime_\kappa =\phi_\kappa$) 
(by which we imply $\phi_\kappa$, which is the actual physically
meaningful phase, since $\varphi$ can be absorbed into 
$\phi^\prime_\kappa$ by a field re-definition) can lead to a 
relatively improved consistency with the experimental data.

The reason for choosing $\phi^\prime_\kappa$ as the only variable
phase, while setting $\theta_{1/2}$, $\theta_{\tilde{f}}$ and
$\phi^\prime_\lambda$ to $0^\circ$, is that it is virtually
unconstrained by the measurements of fermionic 
EDMs~\cite{Cheung:2010ba,*Cheung:2011wn,Graf:2012hh}. Furthermore,
our aim here is to analyse the scenarios with a generic CPV
phase and compare them with the rNMSSM limit rather than 
measuring the effect of any of the individual phases. Note however
that, since only the difference $\phi^\prime_\lambda-\phi^\prime_\kappa$
enters the Higgs mass matrix at the tree level, the impact of a
variation in $\phi^\prime_\lambda$ is also quantified by that due to the
variation in $\phi^\prime_\kappa$ at this level. At higher orders
though, a variation in $\phi^\prime_\lambda$ has an impact on the sfermion and 
neutralino/chargino sectors which is independent of $\phi^\prime_\kappa$. 

In our numerical analysis, we used the program
NMSSMCALC-v1.03~\cite{Baglio:2013iia} for computing the Higgs boson 
mass spectrum and decay branching ratios (BRs) for a given model input
point. The public distribution of NMSSMCALC contains two 
separate packages, one for the rNMSSM only and the
other for the cNMSSM. 
%In principle one can use the package for
%the cNMSSM to study also the rNMSSM, by setting all the CPV phases
%to $0^\circ$. However, 
Some supersymmetric corrections to the Higgs boson decay widths are currently
 only available in the rNMSSM and hence are not included in the
 cNMSSM package. For consistency among our rNMSSM
 and cNMSSM results, we therefore set $\phi_\kappa =0^\circ$ in
 the cNMSSM package for the rNMSSM case instead of using the dedicated
 rNMSSM package. Furthermore, using the cNMSSM code also for the
 rNMSSM limit makes it convenient to draw a one-on-one correspondence 
 between the $\phi_\kappa =0^\circ$ case and each of the
 $\phi_\kappa > 0^\circ$ cases in a given scenario. 
 This is because in the cNMSSM package, even in the rNMSSM limit, the
 five neutral Higgs bosons are ordered
 by their masses and not separated on the basis of their CP-identities. 
 Thus, the scenario with mass-degenerate $H_1,\, H_2$, which we will
 henceforth refer to as
 the $\hobs = H_1+H_2$ scenario, takes into account both the $\sim 125$\,GeV
$H_1,\,H_2$ as well as the $\sim 125$\,GeV $H_1,\, A_1$ solutions of 
the rNMSSM without distinguishing between them. If one, conversely, 
uses the rNMSSM package, these two scenarios ought to be considered separately.
%and then 
%merged together for comparing the fits to the data that take into
%account all possible solutions for $\phi_\kappa = 0^\circ$ in a given 
%scenario against those for $\phi_\kappa > 0^\circ$. 
The same is true also for the $\hobs = H_2+H_3$ scenario, 
wherein $H_2,\, H_3$ are mass-degenerate.  
%However, we also performed a test scan for the
% $H_1+H_2$ case using the rNMSSM package in order to confirm that the
% impact of the added corrections on our overall inferences is
% sufficiently small. 

 The program NMSSMCALC allows one the option to include only 
 the complete 1-loop contributions in the Higgs mass matrix or to
 add also the 2-loop $\mathcal{O}(\alpha_t\alpha_s)$
 corrections to it. For our analysis, in order for better theoretical
 precision, we evaluated the Higgs boson
 masses at the 2-loop level. 
In the NMSSMCALC input, one also needs to choose between 
the modified dimensional regularisation ($\overline{\rm DR}$) and on-shell
 renormalisation schemes for calculating contributions from the top/stop
 sector in the program. We opted for the $\overline{\rm DR}$ scheme for
 each scenario. Note though that further inclusion of 
 $\mathcal{O}(\alpha_b \alpha_s)$, $\mathcal{O}(\alpha_t + \alpha_b +
 \alpha_\tau)^2$ and the recently calculated NMSSM-specific
 $\mathcal{O}(\alpha_\lambda+\alpha_\kappa)^2$ 2-loop
 corrections~\cite{Goodsell:2014pla} in NMSSMCALC may 
have a non-negligible impact on the Higgs 
boson masses and observables~\cite{private}. We, however, maintain that
 such contributions will only result in a slight shifting of the parameter
 configurations yielding solutions of our interest here, but our
 overall results and conclusions should still remain valid.

We performed six sets of scans of the cNMSSM parameter space by
linking NMSSMCALC with the 
MultiNest-v2.18~\cite{Feroz:2007kg,*Feroz:2008xx,*Feroz:2013hea} 
package. MultiNest performs a multimodal sampling of a theoretical
model's parameter
space based on Bayesian evidence estimation. However, we use this
package not for drawing Bayesian inferences about the various NMSSM scenarios 
considered but simply to avoid a completely random sampling of the
9-dimensional model parameter space. In the program, we therefore defined a
Gaussian likelihood function for the $\hobs$ in a given
scan, assuming the experimental measurement of its mass to be 125\,GeV 
and allowing upto $\pm2$\,GeV error in its model prediction. 
We set the enlargement
factor reduction parameter to 0.3 and the evidence tolerance factor to
a rather small value of 0.2, so that while the package sampled
more concentratedly near the central mass value, a sufficiently large number of
points were collected before the scan converged.  
In each of the first two sets of scans we required $H_1$ to be the $\hobs$. In the
third set we imposed this requirement of consistency with the $\hobs$
mass on $H_2$, in the fourth on $H_3$, in the fifth on both
$H_1,\,H_2$ and in the sixth on
both $H_2,\,H_3$. Each of the six sets further contained five separate
scans corresponding to $\phi_\kappa =0^\circ\,,3^\circ\,,10^\circ\,, 30^\circ$ and
$60^\circ$. 

The scanned ranges of the nine free
parameters (after fixing the phases) of the natural NMSSM, which are
uniform across all its five scenarios considered, are given in
tab.~\ref{tab:params}(a). 
Only large
values of $\lambda$ and $\kappa$ are used in this model (with the
upper cut-off on
them imposed to avoid the Landau pole). Since large
radiative corrections from SUSY sectors are not necessary in the
natural limit of the NMSSM, the parameters $M_0$, $M_{1/2}$ and $A_0$
are not required to take too large values. Note that while $A_0$ can
in principle be both positive and negative, with a slightly
different impact on the physical mass of the SM-like
Higgs boson for an identical set of other input
parameters in each case, we restricted the scans to its negative
values only, in order to increase the scanning efficiency.

 In the remaining sixth scan, we considered the complementary parameter space
of the NMSSM, with $\lambda$ and $\kappa$ kept to relatively 
smaller (and $\tan\beta$ to larger) values, so as to prevent too
large a singlet-doublet mixing. In fact, for $\lambda,\kappa \to 0$, 
when the singlet sector gets effectively decoupled, $H_1$, which is by default identified
with the $\hobs$, has properties very identical to the lightest
Higgs boson of the MSSM. Since $H_1$ in
such a case does not obtain a maximal tree-level mass that is possible in the
most general model, large radiative corrections are needed from the
SUSY sector. Hence we used slightly
extended ranges of the remaining parameters, which are given in 
tab.~\ref{tab:params}(b). This scenario, which we refer to as the
low-$\lambda$-NMSSM scenario henceforth, has been included in our analysis in
order to compare the inferences made for the natural NMSSM with an
approximate MSSM limit of the model.

\begin{table}[tbp]
\begin{center}
\begin{tabular}{cc}
\subfloat[]{%
\begin{tabular}{|c|c|}
\hline
Parameter & Natural NMSSM range  \\
\hline
$M_0$\,(GeV) 	& 200 -- 2000 \\
$M_{1/2}$\,(GeV)  & 100 -- 1000	\\
$A_0$\, (GeV)  & $-3000$ -- 0\\
$\tan\beta$ 		& 1 -- 8 \\
$\lambda$ 		& 0.4 -- 0.7 \\
$\kappa$ 		& 0.3 -- 0.6 \\
$\mu_{\rm eff}$\,(GeV) 	& 100 -- 300 \\
$A_\lambda$\,(GeV)  	& $-1000$ -- 1000 \\
$A_\kappa$\,(GeV)  	& $-1000$ -- 1000\\
\hline
\end{tabular}
}
&
\subfloat[]{%
\begin{tabular}{|c|c|}
\hline
Parameter & Low-$\lambda$-NMSSM range  \\
\hline
$M_0$\,(GeV) 	& 200 -- 4000 \\
$M_{1/2}$\,(GeV)  & 100 -- 2000	\\
$A_0$\, (GeV)  & $-7000$ -- 0\\
$\tan\beta$ 		& 5 -- 45 \\
$\lambda$ 		& 0.001 -- 0.4 \\
$\kappa$ 		& 0.001 -- 0.3 \\
$\mu_{\rm eff}$\,(GeV) 	& 100 -- 2000 \\
$A_\lambda$\,(GeV)  	& $-1000$ -- 4000 \\
$A_\kappa$\,(GeV)  	& $-4000$ -- 1000\\
\hline
\end{tabular}
}
\end{tabular}
\caption{Ranges of the NMSSM parameters scanned, with fixed
  $\phi_\kappa$, for (a) each $\hobs$ scenario in the natural NMSSM
  and (b) the low-$\lambda$-NMSSM scenario.}
\label{tab:params}
\end{center}
\end{table}
 
Once the scans had completed, we filtered the points
obtained with each by further imposing 
$123\,{\rm GeV}  \leq m_{\hobs} \leq 127\,{\rm GeV}$. Note that in
the $\hobs = H_1+H_2$ and $\hobs = H_2+H_3$ scenarios,
this condition was imposed on $H_2$, since in both these scenarios it
 is typically the Higgs boson with SM-like couplings. The total number of
points, $N_{\rm total}$, remaining after this filter is given in 
tab.~\ref{tab:points} for each scenario considered.
All these points were then tested for consistency with the
LEP and LHC exclusion limits on the other, non-SM-like,
Higgs bosons of the model, using the package
\HiB~\cite{Bechtle:2008jh,*Bechtle:2011sb,*Bechtle:2013gu,*Bechtle:2013wla}.
The points passing the HiggsBounds test were retained as the `good points'
for further analysis, and their number, denoted
by $N_{\rm HB}$, for each scenario is also given in tab.~\ref{tab:points}. 
We point out for later reference that in each of the two $\hobs = H_1$
scenarios as well as in the $\hobs = H_1+H_2$ scenario, the number of
surviving good points (where they are available) 
is very identical across all input values of $\phi_\kappa$, 
implying mutually fairly consistent sample sizes.

\begin{table}[tbp]
\centering\begin{tabular}{|c|c|c|c|c|c|c|}
\hline
 Scenario  & low-$\lambda$ & $H_{\rm obs} = H_1$  &$H_{\rm obs} = H_2$&$H_{\rm obs} = H_3$&$H_{\rm obs} = H_1+H_2$&$H_{\rm obs} = H_2+H_3$\\
\hline
\hline
             \multicolumn{7}{|c|}{$\phi_\kappa=0^\circ$} \\
\hline
$N_{\rm total}$	&  17786 & 15675 & 15072 & 14431 & 26045 & 23736 \\
$N_{\rm HB}$	& 17722  & 13691 & 2904 & 965 & 11878 & 2819 \\
%$N_{\chi^2<69}$  &  69.5 & & & & &\\
\hline
             \multicolumn{7}{|c|}{$\phi_\kappa=3^\circ$} \\
\hline
$N_{\rm total}$	& 17829 & 15775 & 15026 & 14806 & 27199 & 25684 \\
$N_{\rm HB}$	& 17782 & 13885 & 3235 & 2391 & 11863 & 1659 \\
%$N_{\chi^2<69}$  &  69.5 & & & & &\\
\hline
              \multicolumn{7}{|c|}{$\phi_\kappa=10^\circ$} \\
\hline
$N_{\rm total}$	& 17847 & 15784 & 15080 & 14810 & 26735 & 28348 \\
$N_{\rm HB}$	& 17786 & 13866 & 2411 & 2495 & 12607 & 3369 \\
%$N_{\chi^2<69}$  &  69.5 & & & & &\\
\hline
              \multicolumn{7}{|c|}{$\phi_\kappa=30^\circ$} \\
\hline
$N_{\rm total}$	& 17810 & 16256 & 15037 & 14671 & 31719 & 28685 \\
$N_{\rm HB}$	& 17743 & 14725 & 247 & 276 & 13503 & 2012 \\
%$N_{\chi^2<69}$  &  69.5 & & & & &\\
\hline
              \multicolumn{7}{|c|}{$\phi_\kappa=60^\circ$} \\
\hline
$N_{\rm total}$	& 17810 & 0 & 14996 & 14438 & 0 & 30412 \\
$N_{\rm HB}$	& 17743 & 0 & 247 & 2 & 0 & 242 \\
%$N_{\chi^2<69}$  &  69.5 & & & & &\\
\hline
\end{tabular}
\caption{Number of scanned points remaining after imposing the mass
  constraint on $\hobs$ and those passing the HiggsBounds test, for
  each scenario studied. See text for details.}
\label{tab:points}
\end{table}

Next we carried out fits to the $\hobs$ data for the good points using 
the public code \HiS~\cite{Bechtle:2013xfa}. For obtaining these fits, 
HiggsSignals requires, along with
the masses and BRs of each Higgs boson, $H_i$, 
the square of its normalised effective couplings, $(g_{H_iX}/g_{h_{\rm
    SM}X})^2$, to a given SM particle pair
$X$, with $h_{\rm SM}$ being the SM Higgs boson with the same
mass as the $H_i$. Note that when $X$ is a pair of fermions, there is 
a scalar as well as a pseudoscalar normalised coupling for each $H_i$,
both of which need to be passed separately to HiggsSignals. 
All these are then used to calculate the normalised cross sections,
\begin{eqnarray}
\label{eq:muX}
\mu^X_{H_i} \equiv \frac{\sigma(pp \to H_i \to X)}{\sigma(pp \to h_{\rm SM} \to X)}\,,
\end{eqnarray}
corresponding to a given decay channel, $X$, in an approximate way. 
The NMSSMCALC version we used 
did not provide the normalised Higgs boson couplings as an
output. We therefore modified the
code to obtain these couplings for adding them as a dedicated block in
the SLHA input file for HiggsSignals. 

The program HiggsSignals compares the computed $\mu^X_{H_i}$ for each $H_i$ with 
the experimentally measured ones, $\mu^X_{\rm exp}$, for wide ranges of input
Higgs boson masses in a variety of its production and 
decay channels at the LHC and the Tevatron. We used only 
the `peak-centred' method and the `latestresults' observable set in the
program, with the assignment range variable $\Lambda$ set to the default value
of 1. It thus performed a fit to a total of 81 Higgs boson peak
observables (77 from signal strength and 4 from mass measurements), 
from the CMS, ATLAS, CDF and D$\slashed{\rm O}$ collaborations, for a
given model point. We assumed a Gaussian theoretical uncertainty of 2\,GeV in
the masses of the three lightest neutral Higgs bosons of the model. The
default values of the uncertainties in the Higgs boson production cross sections
as well as BRs were retained. Further details
about the fitting procedure can be found in the manual~\cite{Bechtle:2013xfa} of the 
package. The main output of HiggsSignals contains the total $\chi^2$ and
the $p$-value from the fit, given the number of statistical degrees of
freedom, for each model point. Since the aim of this study is a comparison
of various $\hobs$ scenarios rather than
the overall goodness of fit for each, we will
quantify our results only in terms of the $\chi^2$ and ignore the
$p$-value. 

As an observable indication of the presence of more than
one Higgs bosons near 125\,GeV, the double ratios
\begin{eqnarray}
\label{eq:ratios}
D_1=\frac{R^h_{\rm VBF}(\gamma\gamma)/R^h_{gg}(\gamma\gamma)}{R^h_{\rm VBF}(bb)/R^h_{gg}(bb)}\,;~~D_2=\frac{R^h_{\rm VBF}(\gamma\gamma)/R^h_{gg}(\gamma\gamma)}{R^h_{\rm VBF}(WW)/R^h_{gg}(WW)}\,;~~D_3=\frac{R^h_{\rm VBF}(WW)/R^h_{gg}(WW)}{R^h_{\rm VBF}(bb)/R^h_{gg}(bb)}\,,
\end{eqnarray}
were proposed in~\cite{Gunion:2012he}. Each of these ratios should be unity
if the $\hobs$ constitutes of only a single Higgs boson, while the 
contribution of two (or more) Higgs bosons to the $\hobs$ signal 
could result in a deviation of these ratios from 1. In the above expressions, 
$R^h_Y (X) = R^{H_i}_Y (X) + R^{H_j}_Y (X)$, where $H_i$
and $H_j$ are the two mass-degenerate Higgs bosons in a given scenario
and the subscripts VBF and $gg$ imply the vector boson fusion and
the gluon fusion production modes, respectively. $R^{H_i}_Y (X)$ for 
each $H_i$ is defined as
\begin{eqnarray}
\label{eq:RX}
R^{H_i}_Y (X) \equiv \frac{\Gamma(H_i \to Y)}{\Gamma (h_{\rm SM} \to Y)}
\times \frac{{\rm BR}(H_i \to  X)}{{\rm BR}(h_{\rm SM} \to X)}
 = \frac{C^{H_i}_Y  C^{H_i}_X}{\Gamma_{H_i} /\Gamma_{h_{\rm SM}}}\,,
\end{eqnarray}
with $Y$ being the given production mode and, in the last equality, 
$C^{H_i}_{X(Y)} = \Gamma  (H_i \to X(Y)) /\Gamma (h_{\rm SM} \to X(Y))$,
  the normalised partial decay width of $H_i$ into the $X$ ($Y$)
  pair.\footnote{Note that eq.~(\ref{eq:RX}) assumes that the
    $h_{\rm SM}$-normalised production cross sections
    for the $Y={\rm VBF}$ and $gg$ processes can be approximated by the
    normalised partial decay widths of $H_i$ in the $VV$ and $gg$ decay
    channels, respectively.} 
$\Gamma_{H_i}$ and $\Gamma_{h_{\rm SM}}$ are the total
decay widths of $H_i$ and $h_{\rm SM}$, respectively.

We also evaluated the ratios $D_1$, $D_2$ and $D_3$ for the
points which give reasonably good fits to the data (to be defined later)
in the scenarios with two mass-degenerate Higgs bosons. 
For this purpose, $R^{H_i}_Y (X)$ for each $H_i$ was calculated
by fixing $\Gamma_{h_{\rm SM}}$ in eq.~(\ref{eq:RX}) to
$4.105\times 10^{-3}$\,GeV, which is the value given by the program
HDECAY~\cite{Djouadi:1997yw} for a 125\,GeV $h_{\rm SM}$. 
A change of $\pm 2$\,GeV in the mass of $h_{\rm SM}$ 
 has only a marginal affect on this width, which we ignore. 
For calculating the $\Gamma_{h_{\rm SM}}$ with HDECAY, care was taken
that all the partial decay widths of $h_{\rm SM}$ were evaluated at
the same perturbative order as that implemented in NMSSMCALC for computing
$\Gamma_{H_i}$. Moreover, $C^{H_i}_Y$ is simply the squared normalised
coupling of $H_i$ to a vector boson, $V$, pair for the VBF production mode and 
to a gluon pair for the $gg$ mode. Similarly, $C^{H_i}_X$ implies the
$H_iVV$ and $H_i\gamma\gamma$ normalised couplings squared, respectively, for $X=WW$
and $\gamma\gamma$. All these couplings are thus the same ones obtained from 
NMSSMCALC for passing to HiggsSignals. In the case of $X=b\bar{b}$,
though, there is a scalar and  pseudoscalar coupling for each $H_i$,
as noted above. For this reason, $C^{H_i}_{b\bar{b}}$'s were
calculated using the actual $\Gamma (H_i \to b\bar{b})$ from  
the NMSSMCALC output for a given model point and the 
$\Gamma (h_{\rm SM} \to b\bar{b})$ obtained from HDECAY 
for $m_{h_{\rm SM}}=125$\,GeV. 

\section{\label{sec:results} Results and discussion}

\begin{figure}[tbp]
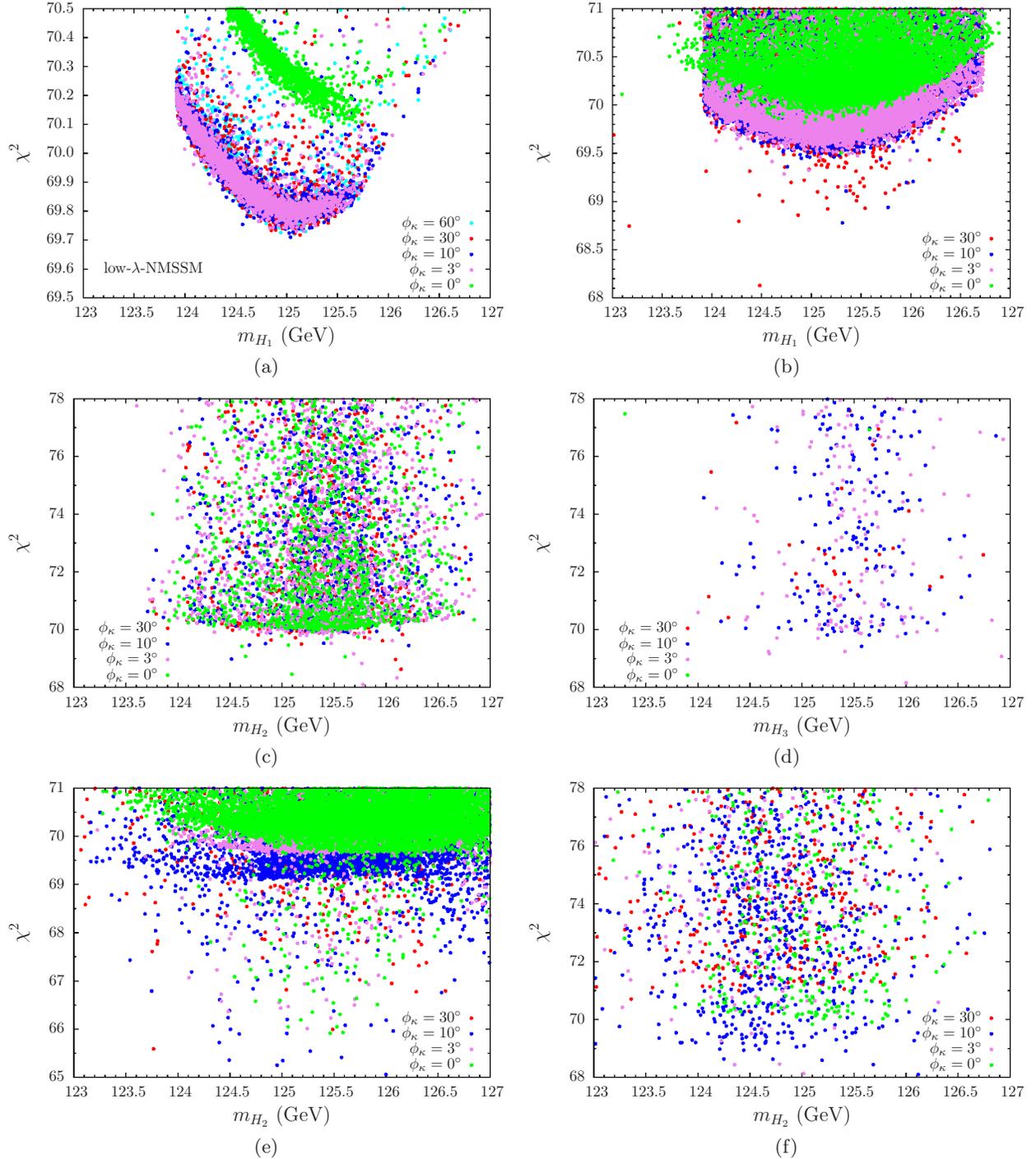

\centering
\begin{tabular}{cc}
%\hspace{-0.1cm}%
\hspace{-0.1cm}%
\subfloat[]{%
\resizebox{0.5\textwidth}{!}{\input{h1_ext.tex}}
} &
\hspace{-0.4cm}%
\subfloat[]{%
\resizebox{0.5\textwidth}{!}{\input{h1_2l.tex}}
} \\

\hspace{-0.1cm}%
\subfloat[]{%
\resizebox{0.5\textwidth}{!}{\input{h2_2l.tex}}
} &
\hspace{-0.4cm}%
\subfloat[]{%
\resizebox{0.5\textwidth}{!}{\input{h3_2l.tex}}
} \\

\hspace{-0.1cm}%
\subfloat[]{%
\resizebox{0.5\textwidth}{!}{\input{h1h2_2l.tex}}
} &
\hspace{-0.4cm}%
\subfloat[]{%
\resizebox{0.5\textwidth}{!}{\input{h2h3_2l.tex}}
} \\
\end{tabular}
\caption[]{Total $\chi^2$ as a function of: (a,\,b) $m_{H_1}$ when only 
  $H_1$ is assumed to be the $H_{\rm obs}$, for two different sets of 
scanned ranges of the parameter space (see text for details); (c) $m_{H_2}$ when only
  $H_2$ is the $H_{\rm obs}$; (d) $m_{H_3}$ when only $H_3$ is the
  $H_{\rm obs}$; (e) $m_{H_2}$ when both $H_1$ and $H_2$ lie near
  125\,GeV and (f) $m_{H_2}$ when both $H_2$ and $H_3$
  lie near 125\,GeV.}
\label{fig:hxfit}
\end{figure}

In fig.~\ref{fig:hxfit} we show the total $\chi^2$ obtained for the
points from our scans for the various $\hobs$ scenarios considered.
%wherein, alternatively, only 
%one of the three lightest neutral Higgs bosons is assumed to play 
%the role of the $\hobs$, against its mass. 
The green points in the figure 
correspond to $\phi_\kappa=0^\circ$, violet to $\phi_\kappa=3^\circ$, 
blue to $\phi_\kappa=10^\circ$, red to $\phi_\kappa=30^\circ$ and
cyan to $\phi_\kappa=60^\circ$. 
For the scenarios in which only 
one of the three lightest neutral Higgs bosons is assumed to be
the $\hobs$, we have made sure that
%, for all the points shown in each panel of the figure, 
 the difference between the mass of $\hobs$ and that of each additional Higgs
 boson nearest to it is always larger than 2.5\,GeV. 
%the $\hobs$ satisfies two conditions: (1) its
% mass is larger (smaller) than 127\,GeV (123\,GeV) and (2) the
% difference between its mass and that of the $\hobs$ 
%The reason for the second condition is to reject those points
 %for which an additional neutral Higgs boson satisfies the first condition
 %only marginally, so that it may still have a mass very close
 %to the $\hobs$ mass. 
The lower cut-off in $\chi^2$ in each panel, 
in this figure and in those that follow, varies depending on
the minimum value obtained in the corresponding scenario. 
The upper cut-off in $\chi^2$ for each scenario is chosen so as to include as
many points in the corresponding figures as possible without the 
$\chi^2$ getting more than 10 units larger than the minimum obtained
in that scenario (given that there are 9 statistical degrees of freedom). 

Fig.~\ref{fig:hxfit}(a) corresponds to the low-$\lambda$-NMSSM
scenario. One notices in the figure that for $\phi_\kappa = 0^\circ$
the $\chi^2_{\rm min}$ lies very close to 70, and is thus almost
identical to $\chi^2_{\rm min} = 69.96$ that is given by
HiggsSignals for a SM Higgs boson at a mass of 125.1\,GeV, with the 
same settings as used by us. The input parameters 
(with the exception of $M_0$, $M_{1/2}$ and
$A_0$, which can be adjusted with much more freedom) and the masses of
the three lightest Higgs bosons are given in tab.~\ref{tab:details}. The
negligibly small difference in the $\chi^2_{\rm min}$ value obtained
for the $h_{\rm SM}$ and for the CP-conserving low-$\lambda$-NMSSM
results from the fact that $\lambda$ for the corresponding point in
the latter is non-vanishing, as seen in the 
table, so that the singlet sector is not completely decoupled and an exact
MSSM-limit is not reached. One can notice in the figure and the
table a slightly lower value of
$\chi^2_{\rm min}$ obtained for the sets of points corresponding to non-zero
$\phi_\kappa$ values. However, $\lambda$ for all these points is even larger than
in the CP-conserving limit. Note also that, for all $\phi_\kappa$, most of the points
give $\Delta \chi^2 \leq 1$.

\begin{table}[tbp]
\centering\begin{tabular}{|c|c|c|c|c|c|c|}
\hline
 Scenario  & low-$\lambda$& $H_{\rm obs} = H_1$  &$H_{\rm obs} = H_2$&$H_{\rm obs} = H_3$&$H_{\rm obs} = H_1+H_2$&$H_{\rm obs} = H_2+H_3$\\
\hline
\hline
             \multicolumn{7}{|c|}{$\phi_\kappa=0^\circ$} \\
\hline
$\chi^2_{\rm min}$   & 70.1 & 69.5 & 68.5 & 76.9 & 65.9 & 69.8 \\
\hline
$\lambda$	& 0.046 & 0.582 & 0.653 & 0.48 & 0.597 & 0.597 \\
$\kappa$	& 0.213 & 0.43 & 0.511 & 0.305 & 0.302 & 0.327 \\
$\tan\beta$	& 17.65 & 1.66 & 3.6 & 6.98 & 2.39 & 2.07 \\
$A_\lambda$	& 853.6 & 226.8 & 609.7 & 680.7 & 540.0 & 179.3 \\
$A_\kappa$	& $-2352$ & $-741.4$ & $-666.0$ & 14.05 & $-479.3 $ & $-3.95$\\
$\mu_{\rm eff}$   & 130.0 & 281.5 & 243.7 & 102.6 & 285.2 & 112.3 \\
\hline
$m_{H_1}$	& 125.4 & 125.3 & 122.1 & 66.8 & 123.3 & 115.1 \\
$m_{H_2}$	& 162.8 & 142.1 & 125.1 & 121.0 & 125.5 & 125.1 \\
$m_{H_3}$ & 1828 & 510.6 & 618.5 & 125.7 & 730.0 & 126.6 \\
\hline
             \multicolumn{7}{|c|}{$\phi_\kappa=3^\circ$} \\
\hline
$\chi^2_{\rm min}$   & 69.7 & 69.2 & 68.1 & 68.2 & 66.0 & 68.1 \\
\hline
$\lambda$	& 0.184 & 0.639 & 0.588 & 0.662 & 0.631 & 0.636 \\
$\kappa$	& 0.291 & 0.523 & 0.39 & 0.349 & 0.373 & 0.318 \\
$\tan\beta$	& 29.6 & 1.81 & 2.61 & 4.24 & 1.61 & 6.45 \\
$A_\lambda$	& 2175 & 162.5 & 459.6 & 425.6 & 222.0 & 848.6 \\
$A_\kappa$	& $-236.7$ & $-595.1$ & $-597.6$ & $-12.03$ & $345.2$ & $-19.4$ \\
$\mu_{\rm eff}$   & 177.9 & 218.8 & 260.5 & 110.1 & 196.4 & 127.4 \\
\hline
$m_{H_1}$	& 125.1 & 125.3 & 122.5 & 97.2 & 123.4 & 105.1 \\
$m_{H_2}$	& 444.9 & 141.7 & 125.8 & 122.3 & 125.2 & 125.0 \\
$m_{H_3}$ & 496.1 & 405.5 & 563.6 & 126.0 & 366.3 & 127.2 \\
\hline
              \multicolumn{7}{|c|}{$\phi_\kappa=10^\circ$} \\
\hline
$\chi^2_{\rm min}$   & 69.7 & 68.8 & 69.0 & 69.4 & 65.1 & 68.1\\
\hline
$\lambda$	& 0.138 & 0.68 & 056 & 0.692 & 0.688 & 0.585 \\
$\kappa$	& 0.219 & 0.409 & 0.345 & 0.338 & 0.361 & 0.306 \\
$\tan\beta$	& 16.7 & 1.85 & 1.91 & 4.88 & 1.98 & 7.55 \\
$A_\lambda$	& 1379 & 291.6 & 347.8 & 557.0 & 390.8 & 972.6  \\
$A_\kappa$	& $-623.8$ & $-476.1$ & $-567.8$ & 12.7 & $-435.1$ & $-30.62$ \\
$\mu_{\rm eff}$   & 133.5 & 251.0 & 266.9 & 124.3 & 254.0 & 136.7 \\
\hline
$m_{H_1}$	& 125.0 & 125.3 & 120.3 & 106.4 & 123.6 & 118.7 \\
$m_{H_2}$	& 212.2 & 140.5 & 124.5 & 111.6 & 126.0 & 126.1 \\
$m_{H_3}$ & 631.6 & 482.5 & 541.8 & 125.6 & 440.1 & 127.4 \\
\hline
              \multicolumn{7}{|c|}{$\phi_\kappa=30^\circ$} \\
\hline
$\chi^2_{\rm min}$   & 69.7 & 68.1 & 68.6 & 70.4 & 65.6 & 70.2\\
\hline
$\lambda$	& 0.136 & 0.648 & 0.679 & 0.537 & 0.624 & 0.481 \\
$\kappa$	& 0.219 & 0.319 & 0.586 & 0.303 & 0.388 & 0.311 \\
$\tan\beta$	& 29.4 & 2.2 & 2.13 & 6.55 & 2.10 & 7.67 \\
$A_\lambda$	& 3515 & 570.1 & 295.0 & 702.2 & 345.7 & 796.5 \\
$A_\kappa$	& $-781.0$ & $-398.4$ & $-590.7$ & 7.07 & $330.5$ & $-23.22$ \\
$\mu_{\rm eff}$   & 170.8 & 288.5 & 227.8 & 112.6 & 209.1 & 110.0 \\
\hline
$m_{H_1}$	& 125.1 & 124.5 & 123.1 & 86.5 & 121.6 & 107.1 \\
$m_{H_2}$	& 234.3 & 127.4 & 126.1 & 116.8 & 123.8 & 124.7 \\
$m_{H_3}$ & 857.7 & 462.4 & 507.8 & 124.3 & 405.8 & 125.8 \\
\hline
\end{tabular}
\caption{Input parameters and Higgs boson masses corresponding to the
  points giving the lowest $\chi^2$ for all $\phi_\kappa$ cases in
  each of the $\hobs$ scenarios considered.}
\label{tab:details}
\end{table}

In fig.~\ref{fig:hxfit}(b), which corresponds to the $\hobs = H_1$
scenario in the natural NMSSM, we see that there is a large
 concentration of points above a $\chi^2$ value which is very
 similar to the $\chi^2_{\rm min}$ seen in the adjacent
 fig.~\ref{fig:hxfit}(a), for each corresponding $\phi_\kappa$. 
%For these points, the lowest $\chi^2$ obtainable
 %for a given $\hobs$ mass seems to be smaller for larger
% $\phi_\kappa$. 
%Even for $\phi_\kappa = 30^\circ$, $\hobs$ struggles to reach a mass above 125\,GeV. 
For non-zero $\phi_\kappa$ though, one also sees a few scattered
points with $\chi^2$ lower than that for any of the points in the
 high concentration region. %Among these scattered points, for
 %$\phi_\kappa= 0^\circ$ (rNMSSM limit) there is only a single isolated
 %point which has a $\chi^2$ just below 77. For non-zero $\phi_\kappa$ 
The overall lowest $\chi^2$ lies very close to 68, for $\phi_\kappa =
30^\circ$, with the mass of $\hobs$ for the corresponding point lying 
at 124.5\,GeV. However, according to tab.~\ref{tab:details},
 the mass of $H_2$ for this point is within 3\,GeV of that of
 $H_1$. It is therefore very likely that the relatively better fit for
 this particular point is a result of the assignment of $H_2$ instead
 of or along with $H_1$ to some of the observables, especially when their
 experimental mass resolution is relatively poor. This possibility,
 which implies that our assumption of two 
Higgs bosons being individually irresolvable if their masses lie within
 2.5\,GeV of each other is rather robust, will be discussed further
 later. For $\phi_\kappa = 60^\circ$ none of the
 points obtained in the scan for this scenario had $H_1$ heavier than 123\,GeV. 

In the $\hobs = H_2$ scenario, a much smaller number of points was
passed by HiggsBounds compared to the $\hobs = H_1$ scenario, 
as seen in Fig.~\ref{fig:hxfit}(c), but the $\chi^2_{\rm min}$ is equally
low for most $\phi_\kappa$ here, including $0^\circ$.
%But at the same time a slightly lower overall $\chi^2$ is
%also seen, which corresponds to $\phi_\kappa = 30^\circ$ here. 
%Generally though, there are more points with relatively low $\chi^2$
%for smaller phases than for $\phi_\kappa = 30^\circ$.
%In this scenario, a $\chi^2$ almost equally low is obtained 
%for ot values of $\phi_\kappa$, including $0^\circ$. 
Only for $\phi_\kappa=60^\circ$, while plenty of
points with $m_{H_2} \approx 125$\,GeV were obtained in the
scan, the $\chi^2$ for them is never low enough to appear in the
figure. Once again, in tab.~\ref{tab:details} one can see that, for the
points giving the lowest $\chi^2$ for each $\phi_\kappa$ in this
scenario, $H_1$ always lies within 3-4\,GeV of $H_2$. Hence the
slightly better fit for this point is again made possible by a
contribution of $H_1$ to some search channels. In
fig.~\ref{fig:hxfit}(d) for the $\hobs = H_3$ scenario, although 
very few points with $\Delta \chi^2 < 10$ appear in this scenario
compared to the ones above, the $\chi^2_{\rm min}$ is very similar, except for the 
$\phi_\kappa=0^\circ$ case, when it has a fairly high value of around 77.   

In fig.~\ref{fig:hxfit}(e) is shown the total $\chi^2$ for the $\hobs=H_1+H_2$ 
scenario against the $H_2$ mass. One can observe quite a few
similarities between this figure and the fig.~\ref{fig:hxfit}(b) seen above (for the
$\hobs=H_1$ scenario). There is once again a large
concentration of points with $\chi^2 \gtrsim 69$ for all $\phi_\kappa$
except $60^\circ$, and also many scattered points below it. 
Importantly though, there are many
points in this scenario which give a $\chi^2$ lower than 68, which is
the overall lowest value observed for any other scenario here. 
Most of these points, including the one with the overall
lowest $\chi^2$ of $\sim 65$, correspond to $\phi_\kappa = 10^\circ$,
although some points for other $\phi_\kappa$ can
also be noticed. In fig.~\ref{fig:hxfit}(f) one sees a
$\chi^2_{\rm min}$ of 68 for the $\hobs = H_2+H_3$
scenario also but very few points with $\chi^2 < 71$, in contrast with
the $\hobs = H_1$ and $\hobs = H_1 + H_2$ scenarios but similarly
to the $\hobs = H_2$ and $\hobs = H_3$ scenarios. 

From the above discussion, it is clear that certain points, or parameter
space configurations, in the $\hobs = H_1+H_2$
scenario give the best fit to the current experimental Higgs boson
data. A {\it global} $\chi^2_{\rm min}$, i.e., the lowest $\chi^2$
value across all scenarios examined here, of around 65 has been observed for
 $\phi_\kappa = 10^\circ$ in this scenario, with some points corresponding
 to other values of $\phi_\kappa$ also lying within 1 unit of this
 $\chi^2$. None of the points obtained for the other scenarios gives a
 $\chi^2$ lying even within 3 units of this global minimum, despite the
 number of sampled points for the $\hobs = H_1$ scenario being
 typically larger. The reason for a better fit for some points with
 two nearly degenerate Higgs bosons becomes apparent by looking at the detailed
 output of HiggsSignals. In the peak-centred method, HiggsSignals
 assigns to a given observable the Higgs boson with a mass closest to the measured mass
 provided by the experiment. This mass measurement currently ranges between
 124.7\,GeV to 126.0\,GeV. Thus, when a single Higgs boson is assigned
 to all the observables, the $\chi^2$ contribution is large from the observables
 for which the measured mass lies away from the mass of the assigned
 Higgs boson, and the experimental mass resolution is good. 
On the other hand, when two Higgs bosons lie close to each other, the
one assigned to a given observable is the one for which the difference
of the predicted mass from the experimental value is the smallest, so that
the $\chi^2$ contribution from this observable is minimal. This is as long as the
mass of the other Higgs boson nearby lies outside the experimental mass resolution,
otherwise HiggsSignals automatically assigns both the Higgs bosons to
an individual observable if it improves the fit. 

Some caveats are in order here though. A $\Delta \chi^2 \simeq
 3$ is statistically quite insignificant for drawing any concrete
 inferences about the considered scenarios, since the total number of 
 observables and statistical degrees of freedom is quite large. 
 At the same time, the number of points giving $\Delta \chi^2 \leq
 3$ is also fairly small. Moreover, no other experimental
 constraints have been imposed in our analysis, since the publicly available tools for
 testing these are so far not compatible with the cNMSSM. It is
 thus possible that many of the interesting points may have already
 been ruled out by such
 constraints. However, the aim of this study is not to disregard one 
 scenario in favour of another, but to simply show that, given the current
 experimental data, the scenario with two mass-degenerate Higgs bosons
 in the NMSSM provides as good, if not better, a fit as the scenarios
 with a single Higgs boson near 125\,GeV. This alternative possibility
 even points towards a source of CP-violation beyond the SM and, therefore,
 warrants more dedicated analyses as well experimental probes. In
 the following we discuss some other interesting aspects of this scenario. 

\begin{figure}[tbp]
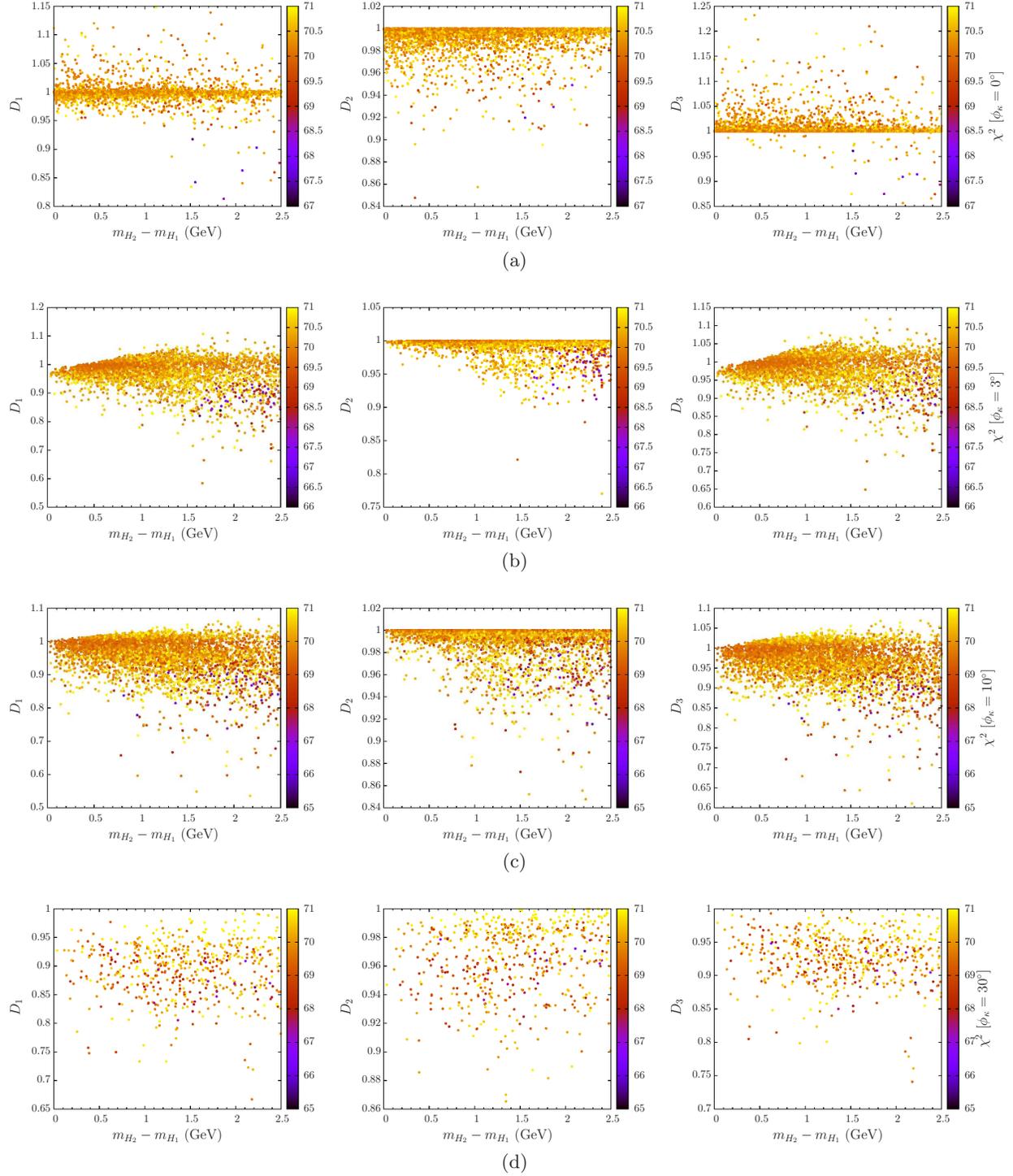

\centering
\begin{tabular}{c}
\subfloat[]{%
\hspace{-0.3cm}%
\resizebox{0.35\textwidth}{!}{\input{h1h2_D1_00deg.tex}}
\hspace{-0.5cm}%
\resizebox{0.35\textwidth}{!}{\input{h1h2_D2_00deg.tex}}
\hspace{-0.5cm}%
\resizebox{0.35\textwidth}{!}{\input{h1h2_D3_00deg.tex}}
} \\

\subfloat[]{%
\hspace{-0.3cm}%
\resizebox{0.35\textwidth}{!}{\input{h1h2_D1_03deg.tex}}
\hspace{-0.5cm}%
\resizebox{0.35\textwidth}{!}{\input{h1h2_D2_03deg.tex}}
\hspace{-0.5cm}%
\resizebox{0.35\textwidth}{!}{\input{h1h2_D3_03deg.tex}}
} \\

\subfloat[]{%
\hspace{-0.3cm}%
\resizebox{0.35\textwidth}{!}{\input{h1h2_D1_10deg.tex}}
\hspace{-0.5cm}%
\resizebox{0.35\textwidth}{!}{\input{h1h2_D2_10deg.tex}}
\hspace{-0.5cm}%
\resizebox{0.35\textwidth}{!}{\input{h1h2_D3_10deg.tex}}
} \\

\subfloat[]{%
\hspace{-0.3cm}%
\resizebox{0.35\textwidth}{!}{\input{h1h2_D1_30deg.tex}}
\hspace{-0.5cm}%
\resizebox{0.35\textwidth}{!}{\input{h1h2_D2_30deg.tex}}
\hspace{-0.5cm}%
\resizebox{0.35\textwidth}{!}{\input{h1h2_D3_30deg.tex}}
} \\

\end{tabular}
\caption[]{The ratios $D_1$, $D_2$ and $D_3$, defined in
  eq.~(\ref{eq:ratios}), as functions of the 
difference between $H_2$ and $H_1$ masses, in the scenario when $\hobs
= H_1+H_2$. 
In (a) $\phi_\kappa$ is set to $0^\circ$, in (b) to
$3^\circ$, in (c) to $10^\circ$ and in (d) to $30^\circ$. The heat map in all the panels
corresponds to the total $\chi^2$.}
\label{fig:h1h2ratios}
\end{figure}

In the left, middle and right panels of fig.~\ref{fig:h1h2ratios} we 
show the ratios $D_1$, $D_2$ and $D_3$, respectively, as functions of the mass
difference, $m_{H_2} - m_{H_1}$, for various $\phi_\kappa$ values  
in the $\hobs = H_1+H_2$ scenario. The heat map 
corresponds to the total $\chi^2$ obtained for the points shown in
each panel. This $\chi^2$ has a uniform upper cut-off of 71 across 
all panels, as in fig.~\ref{fig:hxfit}(e), but its lower cut-off
varies according to the minimum obtained for the $\phi_\kappa$ case
that a given panel corresponds to. According to
fig.~\ref{fig:h1h2ratios}(a), for $\phi_\kappa = 0^\circ$ the three
ratios remain largely close to unity, but deviations up to $15-20$\%
can be seen for some points. 
$D_2$, the ratio dependent on only the bosonic signal
 strengths, only gets smaller than 1 for some points and its
 maximum observed deviation is lower than that of $D_1$ and $D_3$,
 each of which can be both above or below unity.
 Importantly, the points for which a large deviation of each
 ratio from 1 is seen are also generally the ones giving a relatively
 good $\chi^2$ fit to the data.    

A similar trend is seen also for other values of
$\phi_\kappa$. However, deviations of $D_1$ and
$D_2$ from unity by up to $40-50$\% are obtained for $\phi_\kappa =
3^\circ$ (fig.~\ref{fig:h1h2ratios}(b)) and $\phi_\kappa =
10^\circ$ (fig.~\ref{fig:h1h2ratios}(c)), but there are many more
points with significantly large deviations of each of the ratios for the latter
phase compared to the former one. For $\phi_\kappa = 30^\circ$ all the
points appearing in fig.~\ref{fig:h1h2ratios}(d) give $D_1$, 
$D_2$ and $D_3$ smaller than 1 and the overall deviation is
generally smaller than for other non-zero phases but larger than
for the rNMSSM limit. Thus, for this phase, the measured signal 
strengths can provide a clear indication whenever two Higgs bosons
are present near 125\,GeV instead of one. The reason why the
deviations of the three ratios are much smaller overall in the case 
of $\phi_\kappa = 0^\circ$ than for the CPV cases, for points showing
the highest consistency with the data, will become clearer below. 

\begin{figure}[tbp]
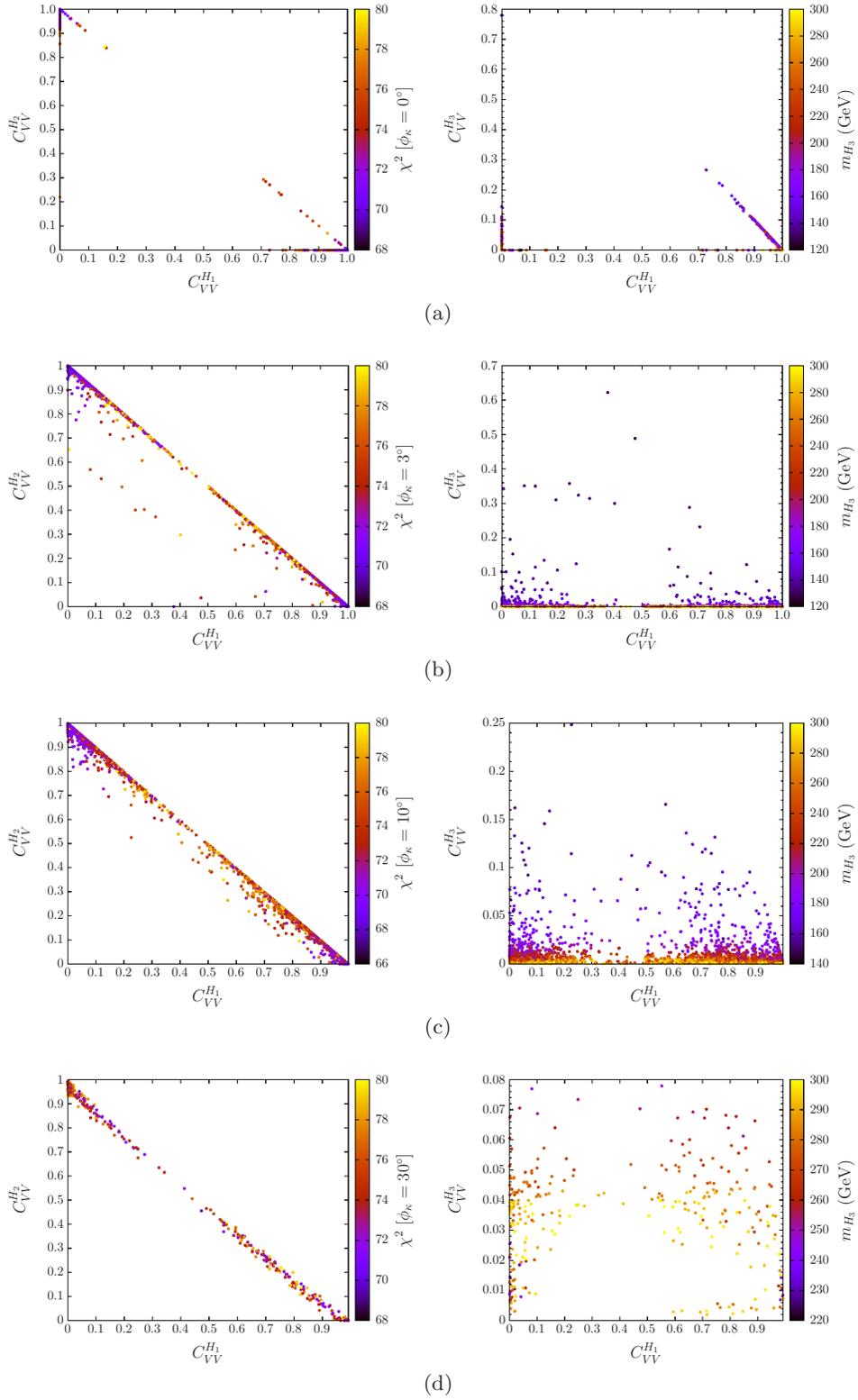

\centering
\begin{tabular}{cc}
\subfloat[]{%
\resizebox{0.38\textwidth}{!}{\input{h1h2_cpl1_00deg.tex}}
\resizebox{0.38\textwidth}{!}{\input{h1h2_cpl2_00deg.tex}}
} \\

\subfloat[]{%
\resizebox{0.38\textwidth}{!}{\input{h1h2_cpl1_03deg.tex}}
\resizebox{0.38\textwidth}{!}{\input{h1h2_cpl2_03deg.tex}}
} \\

\subfloat[]{%
\resizebox{0.38\textwidth}{!}{\input{h1h2_cpl1_10deg.tex}}
\resizebox{0.38\textwidth}{!}{\input{h1h2_cpl2_10deg.tex}}
} \\

\subfloat[]{%
\resizebox{0.38\textwidth}{!}{\input{h1h2_cpl1_30deg.tex}}
\resizebox{0.38\textwidth}{!}{\input{h1h2_cpl2_30deg.tex}}
} \\

\end{tabular}
\caption[]{Squared normalised coupling of $H_1$ to the gauge 
bosons vs. that of $H_2$ (left) and of $H_3$ (right) in the scenario 
when $\hobs = H_1+H_2$, with $\phi_\kappa$ set to (a) $0^\circ$, (b) $3^\circ$,
(c) $10^\circ$ and (d) $30^\circ$. The heat map in the left panels
shows the distribution of $\chi^2$ and in the right panels
that of $m_{H_3}$.}
\label{fig:hxhxcpl}
\end{figure}

As noted earlier, a scenario with two mass-degenerate
Higgs bosons in the cNMSSM entails both the $\hobs
= H_1 + H_2$ and $\hobs = H_1/H_2 + A_1$ possibilities of the rNMSSM. 
Thus it is interesting to see 
%whether one can distinguish between
%points corresponding to either of 
which one of these two possibilities is favoured more by the data, for a given 
$\phi_\kappa$. In the left panels of fig.~\ref{fig:hxhxcpl} we thus
show the squared normalised coupling $C^{H_2}_{VV}$ against
$C^{H_1}_{VV}$, with the heat map corresponding to the total
$\chi^2$. Similarly, in the right panels we have plotted
$C^{H_3}_{VV}$ vs. $C^{H_1}_{VV}$, while the distribution of $m_{H_3}$
is shown by the heat map. For clarity of observation, we
have included in this figure points with a total $\chi^2$ reaching 
up to 80, which is much higher than for the points shown 
in the earlier figures for this scenario. Also we have imposed an upper
cut-off of 300\,GeV on the mass of $H_3$. We expect $C^{H_i}_{VV}$ to 
either vanish when a given $H_i$ is a pure pseudoscalar (in the rNMSSM limit) or
be relatively small when it is pseudoscalar-like (for $\phi_\kappa >
0^\circ$). Note that these couplings satisfy the sum rule~\cite{Cheung:2010ba}
\begin{equation}
\sum_{i=1}^N C^{H_i}_{VV}\simeq 1\,,
\label{eq:sum}
\end{equation}
where $N$ is the total number of neutral Higgs bosons that have a tree-level
coupling to the gauge bosons, i.e., 5 in the cNMSSM and 3 in the
rNMSSM limit.\footnote{Note that since the $h_{\rm SM}$ is a
    hypothetical SM Higgs boson with the
    same mass as a given $H_i$, at the tree level the ratio $C^{H_i}_X$
    in fact corresponds to $(g_{H_iX}/g_{h_{\rm SM}X})^2$ and the
    equality in eq.~(\ref{eq:sum}) is exact. However, since
    $C^{H_i}_X$ have actually been defined here in terms of the
    partial decay widths of $H_i$ in the $X$ channel, which include higher
    order effects also, the sum of $C^{H_i}_X$ may deviate slightly from unity.}
    In the figure we see the above sum rule being satisfied almost completely
by the three lightest neutral Higgs bosons under consideration here, implying that
the remaining two doublet-like Higgs bosons are nearly decoupled.
 
In the case of $\phi_\kappa= 0^\circ$ (i.e., in the rNMSSM limit) 
in the left panel of fig.~\ref{fig:hxhxcpl}(a), we see two distinct 
kinds of points. There are some points lying along the
diagonal, for which $C^{H_1}_{VV}$ and $C^{H_2}_{VV}$ alone
are enough to satisfy the sum rule in eq.~(\ref{eq:sum}). 
It is further evident from the right panel that
$C^{H_3}_{VV}$ for these points is exactly 0. $H_1$ and $H_2$ in these
points should thus be scalars and $H_3$ a pseudoscalar (i.e., $A_1$). 
But for the majority of the points, which lies along either of the
axes, $C^{H_1}_{VV}$ is nearly 1,
implying it is an almost pure doublet-like scalar, while 
$C^{H_2}_{VV}$ is exactly 0, implying it is a pseudoscalar, or vice versa. 
One can then observe in the right panel that for such
points $C^{H_3}_{VV}$, with $H_3$ being the singlet-like scalar, 
is responsible for the sum rule
being satisfied. Thus when the doublet-like scalar, 
whether $H_1$ or $H_2$, has $C^{H_i}_{VV}$ slightly below 1, $C^{H_3}_{VV}$
is slightly above 0. The mixing of the doublet scalar with $H_3$
increases as its mass decreases, as is evident from the heat map in
the right panel of the figure. As a result, the largest $C^{H_3}_{VV}$,
$\sim 0.8$, is seen for the lowest $m_{H_3}$ obtained, which lies
just above the allowed $\hobs$ mass window.

A closer inspection of the heat map in the left
panel of fig.~\ref{fig:hxhxcpl}(a) reveals that the lowest values 
of $\chi^2$ are obtained for points
lying along one of the axes, i.e., when the doublet-like scalar is
nearly mass degenerate with the pseudoscalar. For points along the
diagonal, the $\chi^2$ is in fact always larger than 71. 
This is the reason for the relatively small deviations 
of $D_1$, $D_2$ and $D_3$ from 1 seen
in fig.~\ref{fig:h1h2ratios}(a), where only
points with $\chi^2$ lower than 71 were shown.
For such points, since one of the $H_1$ and $H_2$ is a pure pseudoscalar 
as well as singlet-dominated, its contribution to the combined signal strength in
the $WW$ channel is null and that in the  $\gamma\gamma$ 
and $b\bar{b}$ channels is minimal. 
%This is also the reason why, 
%for $\phi_\kappa = 0^\circ$, the lowest $\chi^2$ seen for the 
%$\hobs = H_1 + H_2$ is almost equal to that for the $\hobs = H_1$ 
%scenario. In short
Therefore, while the presence of $H_1$ and $H_2$ of the
rNMSSM near 125\,GeV may possibly cause $D_1$, $D_2$ and $D_3$ 
to deviate more significantly from 1, the consistency of this scenario with 
the LHC data is worse than that of the $H_1 + A_1$ scenario.   

Fig.~\ref{fig:hxhxcpl}(b) shows that, for
$\phi_\kappa = 3^\circ$, $H_1$ and $H_2$
are almost always scalar-like while $H_3$ is highly
pseudocalar-like with a relatively much smaller
$C^{H_3}_{VV}$ generally. However, due to CP-mixing, $C^{H_3}_{VV}$ can
reach as high as 0.7 or so when the mass of $H_3$ is close to that
of $H_1$ and $H_2$, though this happens for only a few points. 
A very crucial point to note here is that the total
$\chi^2$ in the left panel never
falls below 68, which is due to the cut-off on the allowed upper value
of $m_{H_3}$. This means that the points which give the overall best fit to the
data have a much higher $H_3$ mass, which leads to a much smaller
scalar-pseudoscalar mixing and hence negligible $C^{H_3}_{VV}$.

For the $\phi_\kappa = 10^\circ$ case, illustrated in
fig.~\ref{fig:hxhxcpl}(c), while the maximum $C^{H_3}_{VV}$ obtained
is relatively small and hence $C^{H_1}_{VV}$ and $C^{H_2}_{VV}$ do
not deviate from the diagonal by much in the left panel, there are
many more points, compared to the $\phi_\kappa =
3^\circ$ case above, for which $C^{H_3}_{VV}$ is significant,
according to the right panel. Finally,
for $\phi_\kappa = 30^\circ$, although $C^{H_3}_{VV}$ never completely
vanishes, it also stays smaller overall than it is for other phases. The reason for
this is that the pseudoscalar-like $H_3$ never achieves a mass below
220\,GeV or so, as can be noted from the heat map in the right panel
of fig.~\ref{fig:hxhxcpl}(d). In the left panel one therefore sees
that $C^{H_1}_{VV}$ and $C^{H_2}_{VV}$ always remain very close to the
diagonal. Hence, for non-zero $\phi_\kappa$ the data clearly
favours two scalar-like Higgs bosons near 125\,GeV, instead of a pair
of scalar-like and pseudocalar-like Higgs bosons. 

\section{\label{sec:concl}Conclusions}

In summary, we have tested the consistency of the real and complex NMSSM 
with the latest Higgs boson data from the LHC Run-I and the
Tevatron. In particular, we have
focused on scenarios wherein the resonant peak seen by the
experiments can be explained in terms of
two nearly mass-degenerate Higgs states around 125\,GeV. Such
scenarios have been verified in the rNMSSM previously and have not been
ruled out yet. What we have shown here is that the possibility of such 
dynamics being available in the NMSSM is somewhat enhanced 
%parameter space satisfying 
%(in terms of the percentage of parameter points surviving dedicated model scans in presence of both theoretical and experimental constraints)
if some degree of (explicit) CP-violation is allowed in the Higgs
sector. This can be done by assuming one or more of the Higgs
sector parameters to be complex. By choosing this parameter to be $\kappa$,
one can evade the fermion EDM measurements, which tightly constrain 
the other possibly complex parameters in the Higgs and soft SUSY
sectors of the NMSSM.

In order to achieve the above we have performed 
detailed numerical scans of the parameter space of the cNMSSM 
to obtain various possible
configurations with $\sim 125$\,GeV Higgs boson(s) that also give 
SM-like signal strengths. In these scans we set the phase of
$\kappa$ to five different values, $0^\circ$, $3^\circ$,
$10^\circ$, $30^\circ$ and $60^\circ$. Through a comprehensive analysis
of the points
obtained from these scans, we have then established that
 certain parameter configurations yielding two
Higgs bosons near 125\,GeV 
 are slightly more favoured by the current data compared to 
scenarios with a single $\sim 125$\,GeV Higgs boson.
This statement is even stronger when the two Higgs bosons are CP-mixed states.
For the case of $\phi_\kappa = 10^\circ$ we thus obtained:
 i) the point with the global minimum
 $\chi^2$; ii) more points with $\Delta \chi^2$ lying within 4 units of the
 global minimum $\chi^2$ compared to all
 other scenarios and phases tested; iii) more points with larger deviations of the ratios
$D_1$, $D_2$ and $D_3$ from unity. 

While analysing the aforementioned scenario with two Higgs 
bosons near 125\,GeV, we have made sure that their masses are close enough 
that these two states cannot be distinguished experimentally as separate 
particles. In doing so we have exploited the fact that the
experimental measurements are currently unable
to reconstruct Breit-Wigner resonances, given that the experimental 
resolution in all channels investigated in the Higgs analyses is
significantly larger than the intrinsic Higgs boson widths involved
(so that LHC data actually reproduce Gaussian shapes).
However, (tree-level) interference and
(1-loop) mixing effects become crucial and need to be accounted for
when the (pole) mass difference
between two Higgs states is comparable or smaller that their
individual intrinsic width. While we have ignored such effects here for
points where they can be relevant, which however make up a very tiny
fraction of all the good points from our scans, they are the subject
of a dedicated separate study~\cite{prep}.      

Finally, in our analysis we have used up-to-date sophisticated
computational tools in which state-of-the-art theoretical calculations 
and/or experimental measurements have been implemented, so that 
the solidity of our results is assured. 

\section*{Acknowledgments}

S.~Munir is thankful to Margarete M\"{u}hlleitner for useful 
discussions regarding the cNMSSM and for help with the 
NMSSMCALC program. S.~Moretti is supported in part through
the NExT Institute. S.~Munir is supported by the Korea 
Ministry of Science, ICT and Future Planning, Gyeongsangbuk-Do and 
Pohang City for Independent Junior Research Groups at the Asia Pacific 
Center for Theoretical Physics. 

\bibliographystyle{utphysmcite}	% (uses file "plain.bst")

\bibliography{cNMSSM_deg_refs}

%%%%%%%%%%%%%%%%%%%%%
\end{document}